\documentclass[fleqn,usenatbib]{mnras}

\usepackage{newtxtext,newtxmath,ulem}

\usepackage[T1]{fontenc}

\DeclareRobustCommand{\VAN}[3]{#2}
\let\VANthebibliography\thebibliography
\def\thebibliography{\DeclareRobustCommand{\VAN}[3]{##3}\VANthebibliography}

\usepackage{graphicx}	
\usepackage{amsmath}	

\title[Bayesian forward modelling of cosmic shear]{Bayesian forward modelling of cosmic shear data}

\author[N.\ Porqueres et al.]{
Natalia Porqueres,$^{1}$\thanks{n.porqueres@imperial.ac.uk}
Alan Heavens,$^{1}$
Daniel Mortlock,$^{1,2,3}$
and Guilhem Lavaux$^{4}$
\\
$^{1}$Imperial Centre for Inference and Cosmology (ICIC) \& Astrophysics group, Imperial College, Blackett Laboratory, \\ Prince Consort Road, London SW7 2AZ, UK\\
$^{2}$Department of Mathematics, Imperial College London, London, SW7 2AZ, UK\\
$^{3}$Department of Astronomy, Stockholm University, Albanova, SE-10691 Stockholm, Sweden\\
$^{4}$CNRS \& Sorbonne Universit\'{e}, UMR7095, Institut d'Astrophysique de Paris, F-75014, Paris, France
}

\date{Accepted 21/01/2021. Received 20/01/2021; in original form 13/11/2020}

\pubyear{2020}

\begin{document}
\label{firstpage}
\pagerange{\pageref{firstpage}--\pageref{lastpage}}
\maketitle

\begin{abstract}
    We present a Bayesian hierarchical modelling approach to infer the cosmic matter
    density field, and the lensing and the matter power
    spectra, from cosmic shear data. This method uses a physical
    model of cosmic structure formation to infer physically plausible cosmic
    structures, which accounts for the non-Gaussian features of the
    gravitationally evolved matter distribution and light-cone effects. We test
    and validate our framework with realistic simulated shear data,
    demonstrating that the method recovers the unbiased matter distribution and
    the correct lensing and matter power spectrum. While the cosmology is fixed
    in this test, and the method employs a prior power spectrum, we demonstrate
    that the lensing results are sensitive to the true power spectrum when this
    differs from the prior. In this case, the density field samples are
    generated with a power spectrum that deviates from the prior, and the
    method recovers the true lensing power spectrum. The method also recovers
    the matter power spectrum across the sky, but as currently implemented,
    it cannot determine the radial power since isotropy is not imposed. In summary, our method provides physically plausible inference
    of the dark matter distribution from cosmic shear data, allowing us to
    extract information beyond the two-point statistics and exploiting the full
    information content of the cosmological fields.
\end{abstract}




\section{Introduction}
    
    As light from distant galaxies propagates through the Universe, it is deflected by the gravitational field induced by the large-scale structures. This deflection results in a coherent distortion of observed galaxy images, inducing small changes in the ellipticity of observed galaxies, which is known as cosmic shear. The weak gravitational lensing effect is sensitive to the geometry of the Universe and the growth of cosmic structures, making it a powerful probe to study the matter distribution and the nature of dark matter and dark energy  \citep[see e.g.][for a review]{KilbingerReview}. 
    
    The next-generation surveys like \textit{Euclid} \citep{Euclid2020CosmicShear}, \textit{Roman Space Telescope} \citep{WFIRST} and the Rubin Observatory \citep{LSST} will provide unprecedented precision in cosmic shear measurements, performing wide-field cosmic shear surveys and measuring large and small scales. Harvesting the information from these data sets will present a challenge. Many of the current cosmic shear analyses focus on extracting information from the correlation function or the associated power spectrum \citep{Kitching2011,Heymans2013,Kitching2014,Kitching2015,Alsing16,Kitching2016,Hildebrandt2017,Troxel2018,Hikage2019,Taylor2019}. These analyses capture the two-point statistics, but they do not fully capture the non-Gaussian information encoded in the filamentary features of the matter distribution \citep{Bernardeau1997,Jain1997,Waerbeke1999,Schneider2003,Takada2003,Vafaei2010,Kayo2013}. While some approaches to access the non-Gaussian information are based on measuring high-order correlations \citep{Bernardeau2003,Pen2003,Jarvis2004,Semboloni2011,Waerbeke2013,Fu2014}, peak counts \citep{Jain2000,Dietrich2010,Maturi2011,Marian2012,Pires2012,Cardone2013,Lin2015,Liu2015a,Liu2015b,Kacprzak2016,   Petri2013,Peel2017} or using machine learning \citep{Gupta2018ML}, they rely on summary statistics that do not capture all the information and whose distributions are not well known. 
    
    Capturing the full information content of the large-scale structure requires a field-based approach to infer the matter distribution from observations. \cite{Boehm17} presented a maximum likelihood estimator to reconstruct the matter density field from cosmic shear data, assuming a log-normal distribution for the density. The log-normal distribution reproduces the one- and two-point statistics but fails to reproduce higher-order statistics. \cite{Alsing16, Alsing17} presented a Bayesian hierarchical inference scheme to jointly infer shear maps and the corresponding power spectra, assuming Gaussian statistics of the shear field. From a Bayesian perspective, assuming a Gaussian distribution for the shear field is a well-motivated prior since it constitutes the maximum entropy prior once the mean and covariance are specified. However, more information coming from physics is available, and the Gaussian assumption is sub-optimal. In this work, we address this limitation by including a gravity model in the Bayesian hierarchical model. For this, we build on the Bayesian Origin Reconstruction from Galaxies \citep[BORG,][]{HADES,BORG,BORG-3} framework, which employs a physical description of the dark matter dynamics and allows us to sample from the initial conditions, which are accurately described by Gaussian statistics. With this more complex data model, we get a better representation of the data, and we can extract information beyond the two-point statistics, exploiting the full information content of the shear fields.
    
    One of the main challenges in the analysis of cosmic shear based on estimating the power spectrum is accounting for the masked regions within the survey area \citep[see, e.g.][]{Chon2004,Brown2005,Smith2006}. Our forward modelling approach circumvents these difficulties associated with the survey mask. Although the data do not provide information about the fields in the masked regions, the dynamical model still provides probabilistic information about the shear and density fields that are physically possible in those regions. In our method, the masked regions are treated as pixels with infinite noise, circumventing the need to treat unobserved areas as being cut from the analysis.
    
    \begin{figure}
        \centering
        \hspace{-4em}
    	\includegraphics[width=0.45\columnwidth]{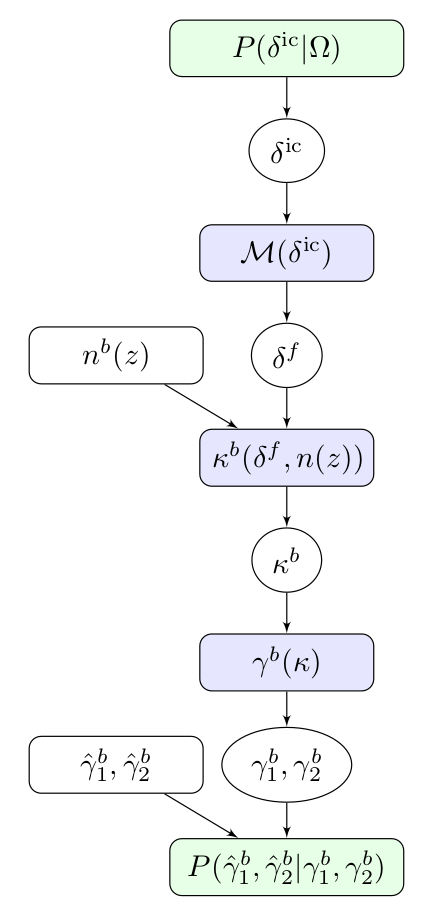}
        \caption{Hierarchical representation of the BORG inference framework for the analysis of cosmic shear data. Primordial fluctuations $\delta_\mathrm{ic}$ encoded in a a set of Fourier modes at $z\approx 1000$ are obtained from the prior $P(\delta_\mathrm{ic}|\Omega)$, where $\Omega$ represents the cosmological parameters. These initial conditions are evolved using the gravity model $\mathcal{M}(\delta^\mathrm{ic})$, which provides the evolved density $\delta_\mathrm{f}$.  The evolved density and the redshift distribution of sources $n^b(z)$ are then used to compute the convergence field for each tomographic bin $b$, $\kappa^b(\delta^f,n^b(z))$. From the convergence, we compute the cosmic shear $\gamma_1^b,\gamma_2^b$ in the flat-sky approximation. $\hat{\gamma}_1^b,\hat{\gamma}_2^b$ indicate the observational data. Purple boxes indicate deterministic transition while green boxes are probability distributions.}
        \label{fig:hierarchical}
    \end{figure}
    
    The paper is organised as follows. Section~\ref{sec:data_model} describes the data model for the cosmic shear and the likelihood. Section~\ref{sec:borg} gives an overview of the Bayesian inference framework, BORG, as required for this work. In Section~\ref{sec:data}, we described the simulated data employed in testing and validating the method. The results are presented in Section~\ref{sec:results}, showing that the method provides unbiased matter density fields. In Section~\ref{sec:prior_test}, we discuss the effect of the prior power spectrum in the results. Finally, Section~\ref{sec:conclusions} summarises the results. 
    
\section{The data model}\label{sec:data_model}

    The effect of weak gravitational lensing on a source can be described by two sky fields: the spin-2 shear, $\gamma$, which describes the distortion in the shape of the image, and the scalar convergence field, $\kappa$, which describes the change in angular size. These two fields are related to the lensing potential, $\phi$, by
    \begin{align}
        \kappa &= \frac{1}{2}\partial \bar{\partial} \phi, \\
        \gamma &= \gamma_1 + i \gamma_2 = \frac{1}{2} \partial \partial \phi,
    \end{align}
    \citep[see e.g.][]{KilbingerReview}, where $\gamma_1$ and $\gamma_2$ are the components of the shear distortion parallel and at $\pi/4$ to the coordinate axes, and $\partial = \partial_x + i \partial_y$ is the complex derivative on the sky, assuming the flat-sky approximation. 
    
    To connect the shear fields to the 3D dark matter distribution, we implemented a line-of-sight integration using the Born approximation, integrating along unperturbed paths. First, we generate convergence fields by integrating along the line-of-sight with the corresponding lensing weights as
    \begin{equation}
        \kappa(\boldsymbol{\theta}) = \frac{3 H_0^2 \Omega_m}{2 c^2} \int^{r_\mathrm{lim}}_0 \frac{rdr}{a(r)} q(r) \delta^f(r\boldsymbol{\theta}, r),
    \end{equation}
    where $\boldsymbol{\theta}$ is the position on the sky, $r$ is the comoving distance, $r_\mathrm{lim}$ is the limiting comoving distance, $\delta^f$ is the final density field and 
    \begin{equation}
        q(r) = \int^{r_\mathrm{lim}}_r dr' n(r') \frac{r'-r}{r'},
    \end{equation}
    where $n(r)$ is the source galaxy distribution. In our discrete implementation, this becomes
    \begin{equation}
        \kappa^b_{mn} =\frac{3 H_0^2 \Omega_m}{2 c^2} \sum\limits_{j=0}^{N_2}  \delta^f_{mnj} \Big[\sum\limits_{s=j}^{N_2} \frac{(r_s - r_j)}{r_s} n^b(r_s) \Delta r_s \Big] \frac{r_j \Delta r_j}{a_j}.
    \end{equation}
    The index $b$ labels the tomographic bin and the subindices $mn$ indicate the 2D pixel on the sky, whose size is chosen to include typically many sources. The sum index $j$ indicates the slice in the radial direction, at a comoving distance $r_j$. $N_2$ is the total number of voxels along the radial axis. The voxels have a length of $\Delta r_j$. $\delta_f$ is the 3D dark matter overdensity at a scale factor $a$. The comoving radial distance $r_s$ indicates the distance to the source plane. The redshift distribution of sources for each tomographic bin is given by $n^b(z_s)$. In this initial proof-of-concept work, we focus on testing the inference and investigating the extent to which the 3D density field, and the 3D matter power spectrum, can be inferred from 2D shear maps. For these tests, we used a simplified scenario, assuming flat-sky and distant observer approximations. In future work, we will drop these approximations and consider the projection effects.
    
    In the flat-sky approximation, we can obtain the shear values from the convergence field. On a flat-sky, the shear and the convergence are related in Fourier space. We, therefore, use a discrete Fourier transform (DFT) to obtain the shear values as
    \begin{equation}
        \gamma^b_{mn}= \mathrm{DFT}^{-1}\left[ \frac{(l_x + i l_y)^2} {l_x^2+l_y^2} \mathrm{DFT}(\kappa^b_{mn})\right],
    \end{equation}
    where $\Vec{l} =( l_x, l_y)$ is the wave-vector written as a complex quantity. Since the convergence is also computed as part of the hierarchical model, this method has the advantage that it can analyse reduced shear,
    \begin{equation}
        g = \frac{\gamma}{1-\kappa},
    \end{equation}
    which, rather than the shear alone, controls the shape distortion.
    
    To analyse cosmic shear observations in our Bayesian framework, we now built a likelihood based on this data model. We assume Gaussian pixel noise for the shear, corresponding to a negative log-likelihood, $\mathcal{L}=-\log P(\hat{\gamma}|\delta^f)$, that can then be written as
    \begin{align}
         \mathcal{L} = 
         \sum\limits_{b}\sum\limits_{mn} \frac{\left[\hat{\gamma}^b_{1,mn}-\gamma^b_{1,mn}(\delta^f)\right]^2 + \left[\hat{\gamma}^b_{2,mn}-\gamma^b_{2,mn}(\delta^f )\right]^2}{2\sigma_b^2},
   \label{eq:likelihood}
   \end{align}
   where $\hat{\gamma} = \hat{\gamma}_1 + i \hat{\gamma}_2$ is the observed data.  This is an estimate of the shear in the pixel, with a variance $\sigma_b^2$, which is determined from the shape noise and number of sources per pixel as $\sigma_\epsilon^2/N_\mathrm{sources}$. We note that even if the ellipticity distribution is not Gaussian, provided many sources contribute to each pixel average the noise will become Gaussian according to the central limit theorem.  An alternative would be to sample from the distribution in another level of the hierarchy, but this would be expensive, so we simplify this stage by using summary statistics of estimated shear and their variance. 
   
   This likelihood is then implemented into the large-scale structure sampler of the BORG framework. The corresponding physical forward modelling approach is illustrated in Fig.~\ref{fig:hierarchical} and proceeds as follows. Using realisations of the three-dimensional field of primordial fluctuations, the dynamical structure formation model evaluates non-linear realisations of the dark matter distribution, accounting for the light-cone effects inherent to deep observations. Using these 3D dark matter field realisations and the data model, BORG predicts shear fields that are compared to the observed data via the likelihood in equation~\eqref{eq:likelihood}. 

    \begin{figure}
    	\includegraphics[width=\columnwidth]{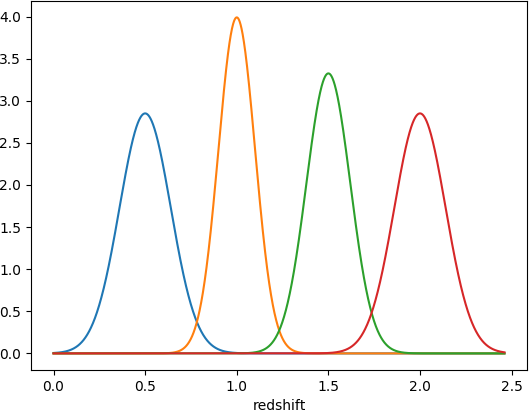}
        \caption{Redshift distributions of sources for the four tomographic bins considered in this analysis.}
        \label{fig:tomobin}
    \end{figure}
    
    \begin{figure}
    	\includegraphics[width=0.9\columnwidth]{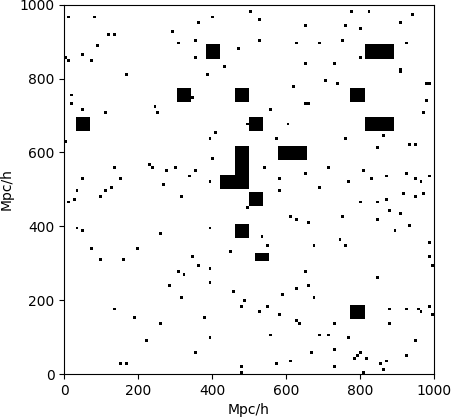}
        \caption{Mask used to generate the mock data. The masked regions, with no contributing sources, are indicated in black. There are two different scales, corresponding to non-observed regions and bright stars.}
        \label{fig:mask}
    \end{figure}

\section{The BORG framework} \label{sec:borg}
    This work extends the previously developed BORG algorithm to analyse the spatial matter distribution underlying cosmic shear observations. In this section, we provide a summary of the algorithm. A more detailed description of the BORG framework can be found in \cite{BORG, JLW15, Lavaux16, Jasche18BorgPM, BORG-3}. 
    
    The BORG framework is a Bayesian inference method aiming at inferring the non-linear spatial dark matter distribution and its dynamics from cosmological data sets. The underlying idea is to fit full dynamical gravitational and structure formation models to observations. By using non-linear structure growth models, the BORG algorithm can exploit the full statistical power of high-order statistics of the matter distribution imprinted by gravitational clustering. This dynamical model links the primordial density fluctuations to the present large-scale structures. Therefore, the forward modelling approach allows the translation of the problem of inferring non-linear matter density fields into the inference of the spatial distribution of the primordial density fluctuations, which are well described by Gaussian statistics \citep{Planck2018nonGaussianity}. The BORG algorithm, therefore, infers the initial matter fluctuations, the dark matter distribution and its dynamical properties from observations. 

    Motivated by inflation theory and observational data, the BORG algorithm employs a Gaussian prior for the initial density contrast at an initial cosmic scale factor of  $a \simeq 10^{-3}$, time for which density perturbations are linearly growing. Initial and evolved density fields are linked by deterministic gravitational evolution mediated by various physics models of structure growth. Specifically, BORG incorporates several physical models based on Lagrangian Perturbation Theory (LPT), fully non-linear particle-mesh models \citep{Jasche18BorgPM}, a model based on spatial COmoving Lagrangian Acceleration framework \citep{Leclercq2020sCOLA}, and a semiclassical analogue to LPT \citep{Porqueres2020PPT}. Any of these dynamical models can be straightforwardly employed within the flexible block sampling illustrated in Fig. \ref{fig:hierarchical}. To test the inference method, in this work, we used LPT to approximately describe the gravitational clustering. However, in a future application of the method to real data sets, we will use the fully non-linear particle-mesh \citep[][]{Jasche18BorgPM} to have a better description of the matter density at small spatial scales, which undergo non-linear dynamics. Though the particle mesh will be more costly, it will still be tractable. \citet{Tassev2013} showed that the LPT begins to show significant deviations at $k>~0.2$ h/Mpc, but using the tCOLA modification of the equation of motion we can push the precision of an LPT-like simulation close to a full $N$-body simulation in a few time-steps, at the field level. Typically, it can be reached in at least as little as ten time-steps to reach 90\% correlation at $k=1h$Mpc$^{-1}$ with a full $N$-body simulation such as one provided by Gadget-2 \citep{Springel2005}.
    
    At its core, the BORG framework employs MCMC techniques. This method allows inference of the full posterior distribution from which we can quantify the uncertainties in our results. However, the inference of the density field typically involves $\mathcal{O}(10^7)$ free parameters, corresponding to the discretised volume elements of the initial conditions. To explore this high-dimensional parameter-space efficiently, the BORG framework uses a Hamiltonian Monte Carlo (HMC) method, which exploits the information in the gradients and adapts to the geometry of the problem. We need, therefore, the adjoint gradient of the data model, which transforms the error vector from the likelihood space to the initial conditions. For the case of weak lensing, we derive this gradient in Appendix \ref{app:gradient}. More details about the HMC and its implementation are described in \cite{jasche2010fast} and \cite{jasche2013bayesian}.

    \begin{figure}
    	\includegraphics[width=\columnwidth]{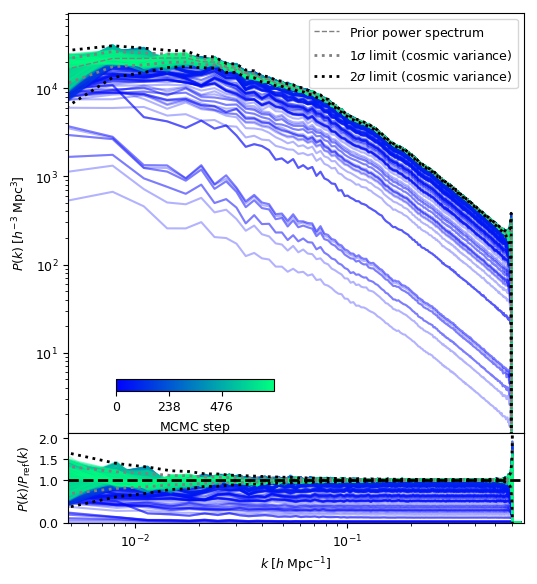}
        \caption{Burn-in of the posterior initial matter power spectra. The colour scale shows the evolution of the matter power spectrum with the number of samples. The dashed lines indicate the underlying power spectrum and the 1- and 2-$\sigma$ cosmic variance limits. The Markov chain is initialised with a Gaussian initial density field scaled by a factor $10^{-3}$ and the amplitudes of the power spectrum systematically drift towards the fiducial values, recovering the true matter power spectrum at the end of the warm-up phase.}
        \label{fig:pk}
    \end{figure}

\section{The mock data} \label{sec:data}
   To test the inference framework, we generated mock observations of cosmic shear with 30 sources per arcmin$^2$ as expected for the Euclid survey, with four tomographic bins and a non-trivial survey mask. In this section, we describe the properties of the synthetic data.  
   
    Mock data are constructed by first generating Gaussian initial conditions on a  Cartesian grid of size 1$h^{-1}$~Gpc $\times 1h^{-1}$~Gpc $\times 4h^{-1}$~Gpc with $128\times 128 \times 256$ voxels. To generate primordial Gaussian density fluctuations we used a cosmological matter power spectrum including the baryonic wiggles calculated according to the prescription provided by \citet[][]{EH98,EH99}. We further assumed a standard $\Lambda$CDM cosmology with the following set of parameters: $\Omega_m = 0.31,\ \Omega_\Lambda = 0.69,\ \Omega_b = 0.049,\ h=0.6711,\ \sigma_8= 0.8,\ n_s = 0.9624$. Here ${\rm H}_0=100 h$~km~s$^{-1}$~Mpc$^{-1}$. 

    To generate realisations of the non-linear density field, we evolve the Gaussian primordial fluctuations via LPT. This involves simulating displacements for $256^2 \times 512$ particles in the LPT simulation, accounting for light-cone effects inherent to deep observations. Final density fields are constructed by estimating densities via the cloud-in-cell scheme from simulated particles on the Cartesian grid. A cosmic shear field is generated by applying the data model described in Sec. \ref{sec:data_model}, assuming the redshift distributions for tomographic bins shown in Fig. \ref{fig:tomobin}. Finally, we added Gaussian pixel-noise to the shear with variance corresponding to $30$ sources per arcmin$^2$, as expected to be obtained from the Euclid survey \citep{EuclidStudyReport}, and with an error on intrinsic ellipticity given by $\sigma_\epsilon = 0.3$ \citep[][]{KilbingerReview}. The total of 30 sources per arcmin$^2$ is then equally distributed between the bins, corresponding to 7.5 galaxies per arcmin$^2$ for each tomographic bin. This corresponds to a signal-to-noise ratio of $S/N = 0.5$. We added a non-trivial survey mask, shown in Fig.~\ref{fig:mask}. Since the data provides no direct information in the masked regions, these are treated as pixels with infinite noise.

    \begin{figure*}
        \centering
            {\includegraphics[width=\hsize,clip=true]{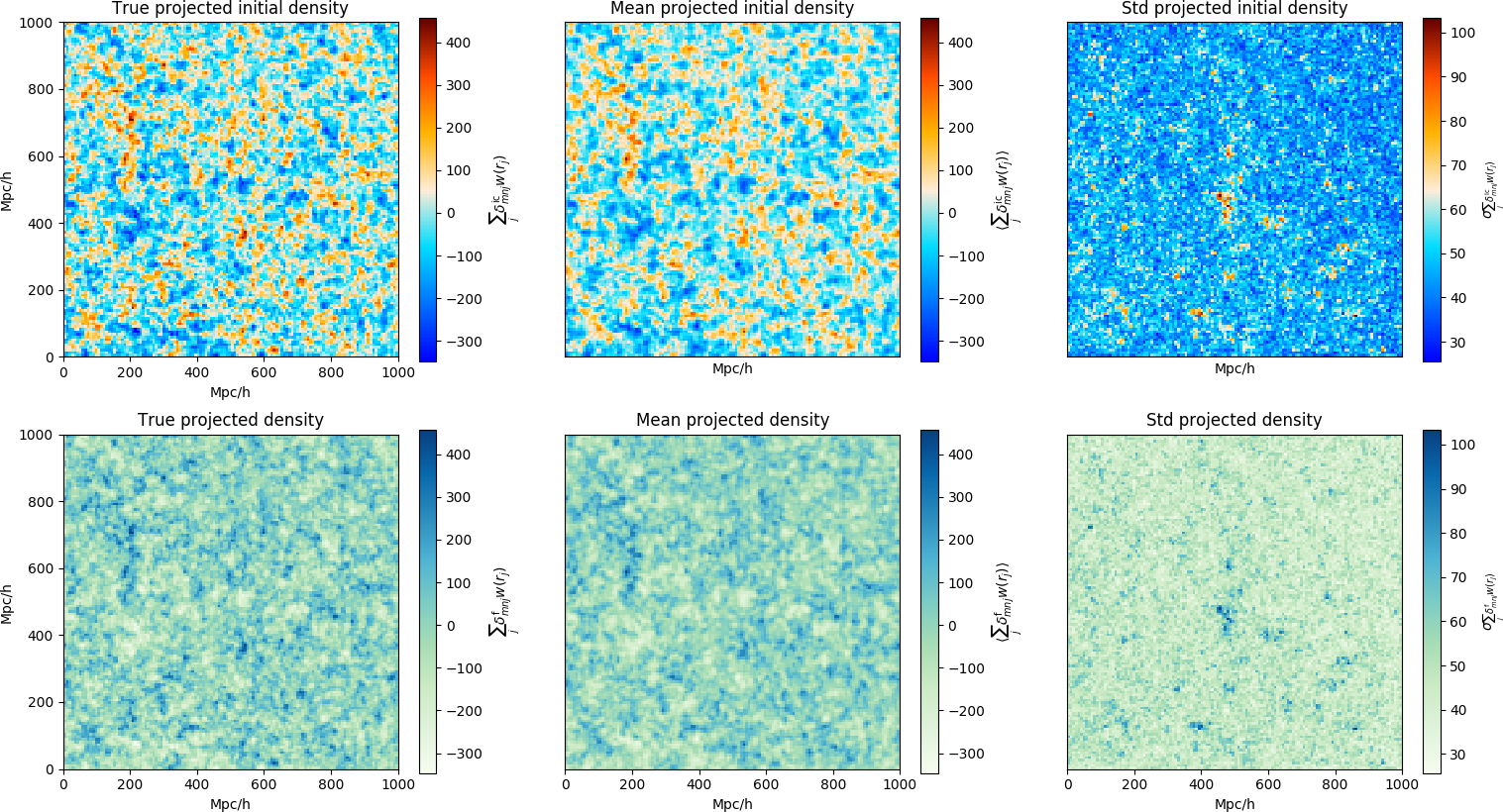}}
        \caption{Projections of the ground truth initial (left upper panel), final density field (left lower panel), inferred ensemble mean initial (middle upper panel) and ensemble mean final (middle-lower panel) density field computed from 500 MCMC samples. Since the information on the radial direction is not very informative, the density fields are projected on the sky, and the different slices of the 3D density field are weighted with the distribution of sources. Comparison between these panels shows that the method recovers the structure of the true projected density field with high accuracy. Right panels show standard deviations of inferred amplitudes of the initial (upper right panel) and final density fields (lower right panel). The regions of high uncertainty correspond to the masked regions, where there are no contributing sources.}
        \label{fig:panels}
    \end{figure*}
    
\section{Results} \label{sec:results}
    Here, we present the results of applying our algorithm to simulated cosmic shear data. We show that our method infers unbiased density fields and corresponding power spectra at all scales considered in this work. We also perform a posterior predictive test for the shear, showing that the inferred densities can explain the data within the noise uncertainty.
    
    \subsection{The warm-up phase of the sampler} \label{sec:warmup}

    In this first Bayesian approach, we keep the cosmology fixed, and specify a prior on the initial power spectrum. However, the power spectrum of the inferred matter distribution is conditioned by the data, and we can use the posterior $P(k)$ as a diagnostic for the effectiveness of the inference since the power spectrum of the posterior samples may differ from the prior. To monitor the initial warm-up phase of the Markov sampler, we follow a similar approach to our previous works \citep{BORG, Foregrounds, Altair, Jasche18BorgPM, RobustLikelihood, PorqueresLya}: we initialised the Markov chain with an over-dispersed state and traced the systematic drift of inferred quantities towards their preferred regions in the parameter space. Specifically, we initialised the Markov chain with a random Gaussian initial density field scaled by a factor $10^{-3}$ and monitored the drift of corresponding posterior power spectra during the warm-up phase. Figure \ref{fig:pk} presents the results of this exercise, showing successive measurements of the posterior power spectrum during the initial warm-up phase. The amplitudes of the posterior power spectrum show a systematic drift towards their fiducial values. By the end of the warm-up phase, the sampler has found an unbiased representation of the initial power spectrum at all Fourier modes considered in this work. Starting the sampler from an over-dispersed state, therefore, provides us with an important diagnostics to test the validity of the sampling algorithm.

\subsection{Inferred density fields} \label{sec:density_results}
    
    As discussed above, our method uses a forward modelling approach, fitting a physical dynamical model to shear data and employs an MCMC sampler to explore the parameter space. This provides the full posterior distribution, from which we draw samples of the initial matter fluctuations. 
    
    Figure \ref{fig:panels}  shows projections of the true fields, and the ensemble mean and variances of inferred three-dimensional fields. The mean and variance are estimated from 500 samples of the posterior distribution (the correlation length is $\approx 80$ samples). To compare the ground truth to the inferred mean density field, we computed the projection of the density fields on the sky since the radial information is not very constraining. In this projection, the different slices of the 3D density field are weighted with the distribution of the lensing sources.  A first visual comparison between ground truth and the inferred ensemble mean initial and final density fields shows that the algorithm correctly recovered the large-scale structures from cosmic shear data. As expected, the mean of the initial density samples exhibits a small degree of smoothing, a feature that is known from the Wiener filtering solution for Gaussian fields and Gaussian prior. 
    
    The right panels of Fig. \ref{fig:panels} show the corresponding standard deviations of the projected densities, which are estimated from the posterior samples. The high uncertainty regions correspond to the masked areas, where there are no contributing sources. While the data do not provide direct information about the density field in these masked regions, the dynamical model still provides probabilistic information about the density fields that are physically plausible in those regions. These results indicate that the method can deal with non-trivial survey masks, and account for the uncertainty in the unobserved areas. The standard deviation of the initial conditions is homogeneous, apart from mask effect, indicating that the dynamical model correctly propagates the information between the primordial matter fluctuations and the final density field.

    \subsection{Posterior predictive tests}
       
    \begin{figure*}
    	\includegraphics[width=\hsize]{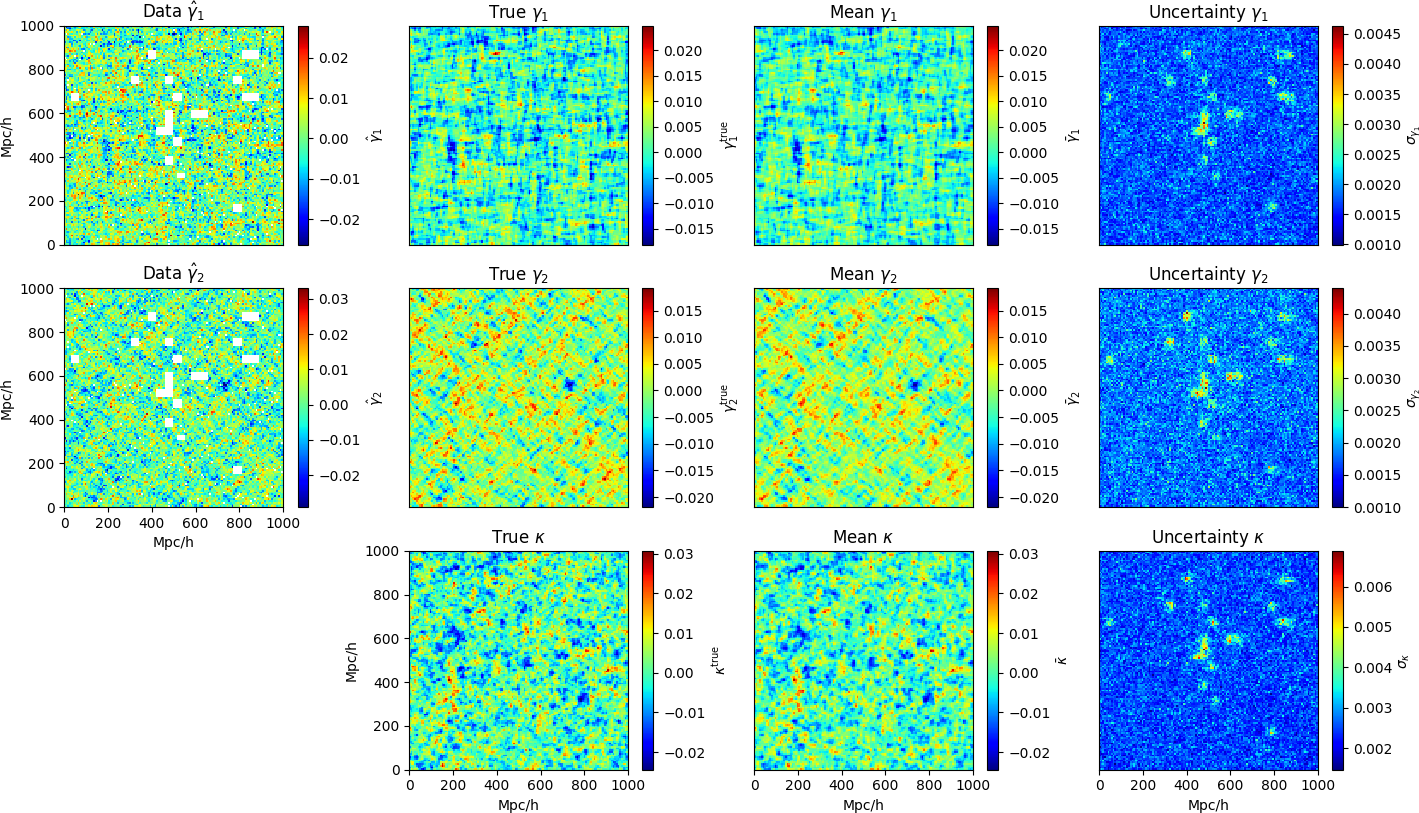}
        \caption{Posterior predicted shear and convergence for one tomographic bin. The left column shows the shear data, including noise and masked regions; the second column shows the true shear and convergence fields, and third and fourth columns show the mean and standard deviation of the posterior-predicted shear and convergence, computed from 500 posterior samples. The method recovers the true cosmic shear correctly. The regions with higher standard deviation correspond to the masked regions.}
        \label{fig:posterior_shear}
    \end{figure*}
    
    \begin{figure}
    	\includegraphics[width=0.95\columnwidth]{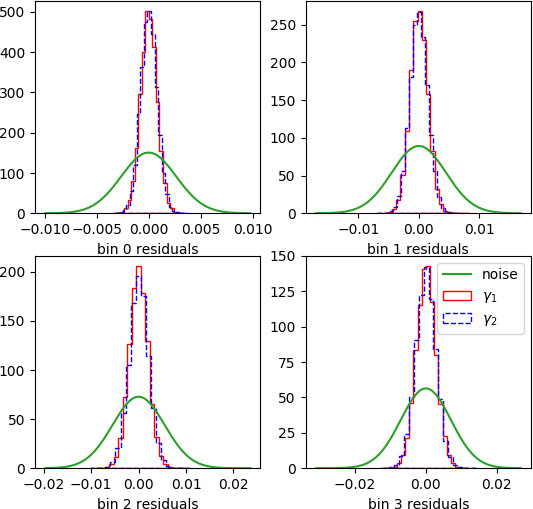}
        \caption{Histogram of the residuals computed as the difference between the posterior predicted shear and the true shear. We note that the true shear does not include the noise. The residuals distribution is narrower than the distribution of pixel noise in the data, indicated in green, showing that the method recovers the true shear at sub-noise level with additional constraints from the cosmological prior.}
        \label{fig:residuals}
    \end{figure}
    
    Posterior predictions allow testing whether the inferred density fields provide accurate explanations of the data \citep[see, e.g.][]{gelmanbda04}. Generally, posterior predictive tests provide good diagnostics about the adequacy of data models in explaining observations and identifying possible systematic problems with the inference. In this section, we predicted the shear and convergence fields as the average computed from 500 posterior samples. 
    
    Figure \ref{fig:posterior_shear} presents the result of this test for one tomographic bin, showing that the posterior predicted shear and convergence recover the features of the true fields. The masked regions show higher standard deviation, indicating that the method can account for the uncertainties in the unobserved areas and provide probabilistic information of the physically plausible shear in those regions. While Fig. \ref{fig:posterior_shear} shows a visual comparison, Fig. \ref{fig:residuals} shows the residuals between the true and the mean posterior-predicted shear fields. The green line in the plot indicates the noise distribution, showing that the distribution of the residuals is narrower than the noise distribution, and, therefore, the inferred quantities can explain the data at sub-noise level, with the additional constraints coming from the cosmological prior.

    \begin{figure*}
    	\includegraphics[width=\hsize]{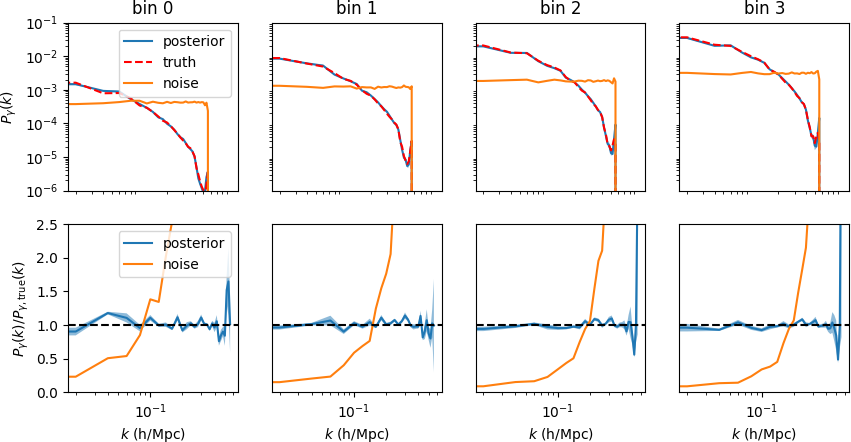}
        \caption{Posterior power spectrum of the shear field compared to the power spectrum of the true shear for each tomographic bin. The posterior power spectrum is the averaged of the power spectrum measured in 500 posterior samples. The orange line shows the noise power spectrum. The bottom plots show the ratio between the posterior and the true power spectrum, showing that the method recovers the true shear power spectrum at all scales.}
        \label{fig:pk_shear}
    \end{figure*}
    
    \begin{figure}
    	\includegraphics[width=\columnwidth]{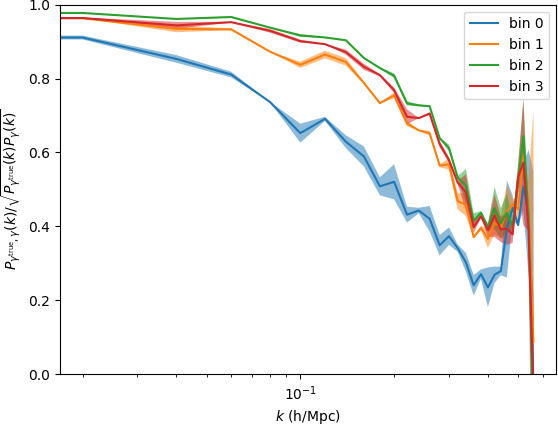}
        \caption{Correlation coefficient between the posterior-predicted and true shear fields, from 500 posterior samples. This shows the expected Wiener-filter-like suppression of power where the noise is high.}
        \label{fig:cross}
    \end{figure}

    Figure \ref{fig:pk_shear} shows the power spectrum of the posterior-predicted shear, measured as the averaged of predicted shear fields from 500 randomly-drawn samples. The predicted power spectra match the true shear power spectrum at all scales. We note that we computed the true power spectrum from the noiseless data as the posterior-predicted shear does not contain noise.  These shear power spectra are pure $E$-mode since they are obtained from the lensing equations under the Born approximation.
    
    Figure \ref{fig:cross} shows the correlation coefficient between the posterior-predicted shear maps and the true shear. This is readily understood, as it is similar to the Wiener solution for the posterior of a statistically homogeneous gaussian field with signal power $S$ and gaussian noise power $N$, \citep[see e.g][] {JH2018}. In this case the mean posterior has power suppressed by $(S^{-1}+N^{-1})N^{-1}$ and this is the correlation coefficient.  As a result, we expect the tomographic bin centred at the lowest redshift (bin 0) to have a lower correlation coefficient because the lensing power at low redshift is smaller.  
    
    \section{Distribution of the convergence field}
    
    \begin{figure}
        \centering
         {\includegraphics[width=\hsize,clip=true]{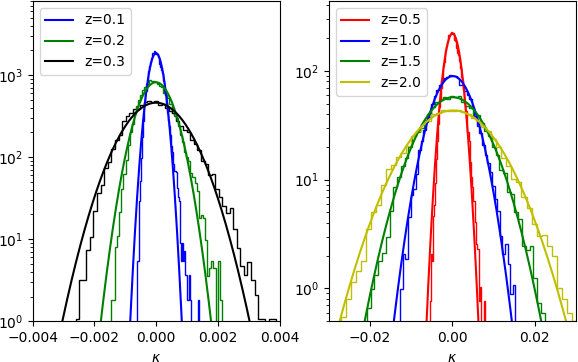}}
         \caption{Distribution of the convergence field for tomographic bins centred at different redshifts. For comparison, a Gaussian with the same mean and variance is plotted on top. The convergence distribution is skewed for tomographic bins centered at $z>0.5$.
         \label{fig:skewness}}
    \end{figure}
    
    Previous Bayesian approaches \citep{Alsing16} rely on a Gaussian prior for the shear data. The Gaussian prior is well-justified when only the mean and variance are known since it is the least informative prior.  It is important to understand that the samples are not Gaussian fields, since they are conditioned on the data, so non-gaussianity in the field may be imposed by the data. However, we can make use of the fact that we have more information available from knowledge of gravitational physics and, for this reason, we include a gravity model in our Bayesian hierarchical model.  The advantage here is that we sample from the initial field, which we know to be Gaussian, so rather than relying on an uninformative prior for the final shear fields, we use the correct Gaussian distribution for the initial conditions.  What we do not yet do in this model is to vary the prior parameters of the power spectrum (as \cite{Alsing16} do), and this will be the subject of future work. As described in section \ref{sec:data_model}, we obtained the shear on a flat-sky from the convergence, which is computed from the non-linear density field. Using our forward model, we have computed the convergence field in a wider range of tomographic bins, centred at different redshifts but with the same bin width ($\sigma=0.1$), to illustrate more clearly how non-Gaussianity in the 1-point distribution evolves. Figure \ref{fig:skewness} shows the distribution of these convergence fields. While the convergence shows a Gaussian distribution for tomographic bins centred at $z>0.5$, it is skewed for tomographic bins at lower redshifts. This indicates that the Gaussian approximation is accurate at large redshift, but it is sub-optimal for low-redshift bins.

\section{Prior test} \label{sec:prior_test}
    
    \begin{figure}
        \centering
         {\includegraphics[width=\hsize,clip=true]{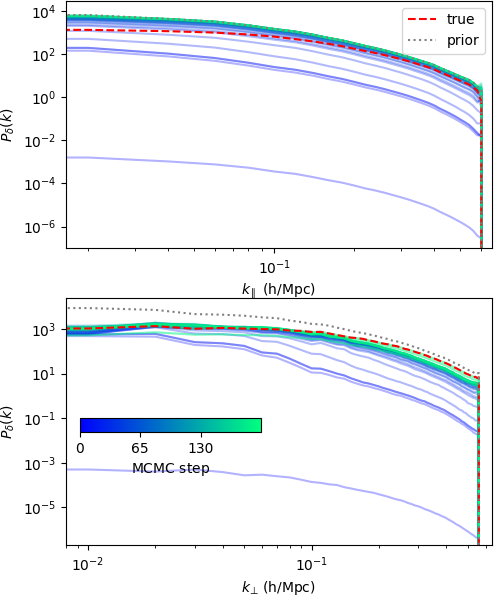}}
         \caption{Burn-in phase of the dark matter power spectra for the prior test. We show the power spectrum for the modes parallel to the line-of-sight, $k_\parallel$ (upper panel), and the perpendicular modes, $k_\bot$ (lower panel), i.e. in the plane of the sky. The dotted line indicates the prior matter power spectrum, and the truth is indicated by the red dashed line. The colour scale indicates the sample number in the Markov chain. After the burn-in phase, the method recovers the correct matter power on the plane of the sky. However, the data is not constraining in the radial direction and the prior dominates. We note that as currently implemented, isotropy is not a requirement. 
         \label{fig:Pk_prior}}
    \end{figure}
    
    \begin{figure}
        \centering
         {\includegraphics[width=\hsize,clip=true]{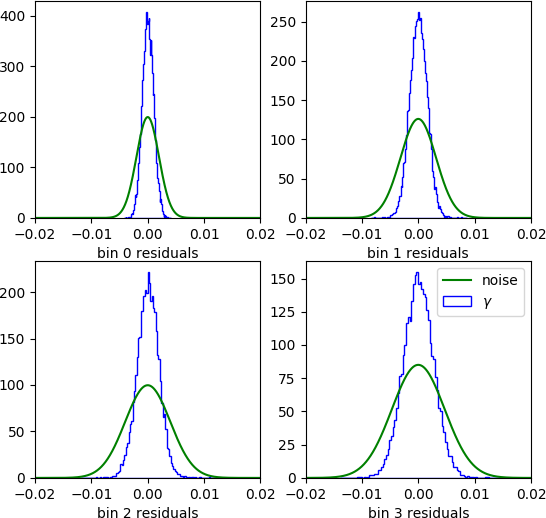}}
         \caption{Histogram of the shear residuals for the prior test. The residuals are computed as the difference between the mean posterior-predicted shear and true shear. The distribution of shear residuals is narrower than the noise distribution, indicated in green.   
        \label{fig:residuals_pk_test}}
    \end{figure}
    
    \begin{figure*}
        \centering
         {\includegraphics[width=\hsize,clip=true]{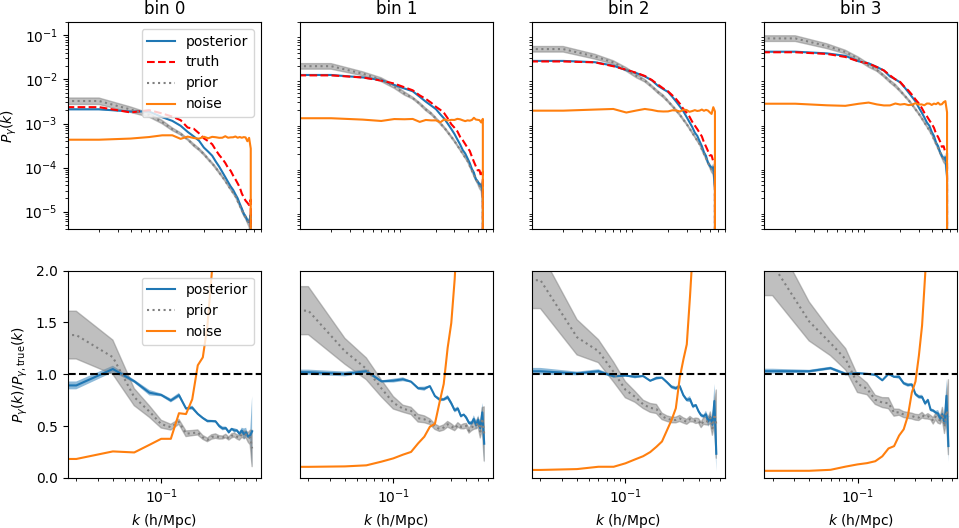}}
         \caption{Posterior power spectrum of the shear field compared to the power spectrum of the true shear for the prior test. The grey dotted line indicates the shear power spectrum computed using the prior cosmology, which differs from the truth. The orange line indicates the power spectrum of the noise. The bottom plots show the ratio between the posterior, prior, and noise and the true power spectrum, showing that the method recovers the true lensing power spectrum where the signal-to-noise is high, but is suppressed in the low S/N regions where the prior is low.
        \label{fig:posterior_pk_shear}}
    \end{figure*}

    As discussed in section \ref{sec:warmup}, in this approach, we keep the cosmology fixed, and specify a prior on the initial power spectrum. Although our method does not sample the power spectrum, the power spectrum of the posterior samples is conditioned by the data. This means that, if the data require it, the posterior power spectrum deviates from the prior, and the density samples have a power spectrum that differs from the prior. To demonstrate that, we tested our method with mock data generated with $h=0.6$,  $\sigma_8=0.55$, $\Omega_m=0.7$ and analyse these data with $h=0.677, \sigma_8=0.8$, $\Omega_m=0.3$, such that $\Omega_mh^2$ changes by a factor 2.  These changes in the cosmological parameters give a different shape and amplitude of the power spectrum, as can be seen in Fig. \ref{fig:Pk_prior}. The resolution, total number of sources per square arcmin, the mask and tomographic bins remain as described in Section \ref{sec:data}.
    
    Figure \ref{fig:Pk_prior} presents the evolution of the power spectrum for this test, showing the burn-in phase of the Markov chain. In the plane of the sky, the posterior power spectrum agrees with the truth, demonstrating that the method is sensitive to the true power spectrum even when this differs from the prior. However, in the line-of-sight direction, the data is not constraining, and the prior dominates. These results are consistent with \cite{Simon2009}. As currently implemented, isotropy is not required, and the method cannot determine the radial power accurately. The recovered matter power spectrum, however, suffices to explain the data, obtaining shear residuals below the noise level, as can be seen in Fig.~\ref{fig:residuals_pk_test}. Figure  \ref{fig:posterior_pk_shear} shows the lensing power spectrum, indicating that the method recovers the correct lensing power at large scales where the noise is low. However, the small scales have lower S/N, and the posterior power spectrum drifts towards the prior. To avoid this prior sensitivity, we need to sample from the power spectrum as well as the field.  This will be the subject of a future paper.
    
    In future work, we will extend our Bayesian hierarchical model to jointly sample the cosmological parameters and the density field, using an approach compatible with the one presented in \cite{Altair}. We expect that this future extension of the method will constrain the matter power spectrum, also in the radial direction, through the assumption of isotropy, but we do not anticipate being able to recover the small-scale radial distribution at the field level, because of the width of the lensing kernel and the distance uncertainties.  Meanwhile, the test presented here shows that the posterior power spectrum is conditioned by the data despite using a fixed cosmology.

\section{Summary and discussion} \label{sec:conclusions}

    We have developed a Bayesian physical forward model to infer the matter density field and primordial fluctuations from cosmic shear data. This framework consists of a Gaussian prior for the primordial fluctuations, a dynamical structure formation model that links the initial conditions and the evolved density field, and a likelihood based on a data model of the cosmic shear in the flat-sky approximation. 
    
    This forward modelling approach allows us to go beyond the common analyses of cosmic shear based on two-point statistics. While many studies of the cosmic shear focus on the power spectrum or the correlation function, the non-linear dynamics of the large-scale structure encode significant information in higher-order statistics associated with the filamentary structure of the cosmic web. Our dynamical forward model reproduces the filamentary matter distribution and, in this way, allows using every data point, rather than relying on summary statistics that do not capture all the information and whose distributions are not well known. By employing a more accurate gravity model, our method also improves over previous Bayesian hierarchical approaches that assumed a Gaussian distribution of the shear field \citep[][]{Alsing16}. 
    
    We have tested our inference method with simulated data with four tomographic bins, a survey mask, and 30 sources per arcmin$^2$ as expected for the Euclid survey. These tests demonstrate that our method recovers the unbiased matter distribution and initial matter power spectrum from cosmic shear data. Posterior predictive tests showed that the inferred quantities are known to the sub-noise level, with additional constraints coming from the cosmological prior. 
    
   Although our framework currently uses a fixed cosmology, we have shown that the method recovers the true power spectrum when this differs from the prior where the signal-to-noise is high. While we do not sample the power spectrum, the posterior power spectrum deviates from the prior if the data require it. To illustrate this, we performed a test using different values of  $H_0$, $\sigma_8$ and $\Omega_m$ to generate the mock data and to analyse them. In this case, the prior power spectrum, therefore, differs from the true power spectrum. This test demonstrated that our method is sensitive to the underlying cosmology, and the power spectrum of the density samples is conditioned by the data, recovering the true matter power spectrum across the sky and the lensing power spectrum. However, in the radial direction, the data is not informative, and the prior dominates since we have not imposed isotropy. In future work, we will extend our Bayesian hierarchical approach to sample the cosmological parameters, through both the geometry and the power spectrum and its growth. We expect that this extension will also constrain the matter power spectrum in the radial direction through the imposition of isotropy, and remove the prior power spectrum sensitivity.
    
    To summarise, this work demonstrates the feasibility of detailed and physically plausible inference of the large-scale structure from cosmic shear data. The proposed approach, therefore, improves the shear data model from previous methods by including a physical description of gravity, providing a better representation of the data and allowing us to extract information beyond the two-point statistics. In future work, we will explore the constraints on cosmology that this approach provides by jointly sampling the initial conditions and the cosmological parameters.

\section*{Acknowledgements}
This work was supported by STFC through Imperial College Astrophysics Consolidated Grant ST/5000372/1. GL acknowledges financial support from the ANR BIG4, under reference ANR-16-CE23-0002. 
This work was carried out within the Aquila Consortium\footnote{\url{https://aquila-consortium.org}}.

\section*{Data Availability}
 The data underlying this article will be shared on reasonable request to the corresponding author.



\bibliographystyle{mnras}
\bibliography{lensing} 




\appendix

\section{Adjoint gradient of the data model} \label{app:gradient}
    The inference of the density field requires inferring the amplitudes of the primordial density at different volume elements of a regular grid, commonly between $128^3$ and $256^3$ volume elements. This implies $10^6$ to $10^7$ free parameters. To explore this high-dimensional parameter space efficiently, the BORG framework employs a Hamiltonian Monte Carlo (HMC) method, which adapts to the geometry of the problem by using the information in the gradients. Therefore, this algorithm requires the derivatives of the forward model. In this appendix, we derive the adjoint gradient of the shear model, which linearly transforms the error vector from the likelihood space to the parameter space of initial conditions.

    More specifically, the HMC relies on the availability of a gradient of the posterior distribution. Therefore, we need to compute the gradient of the log-likelihood with respect to the initial density contrast, $\delta^\mathrm{ic}$.
    
    \begin{equation}
       \frac{\partial \log\mathcal{L}} {\partial \delta^\mathrm{ic}_p} = \sum_b \sum_{mn} \frac{\partial \log\mathcal{L}}{\partial \gamma^b_{mn}} \frac{\partial \gamma^b_{mn}}{\partial \delta^\mathrm{ic}_p}
    \end{equation}

    \begin{align}
       \frac{\partial \log\mathcal{L}} {\partial \delta^\mathrm{ic}_p} &= \sum_b \sum_{mn} 
        \frac{\hat{\gamma}^b_{1,mn} - \gamma^b_{1,mn} }{\sigma^2_{mn}} \frac{\partial \gamma^b_{1,mn}}{\partial \delta^\mathrm{ic}_p} \\ &+
        \frac{\hat{\gamma}^b_{2,mn} - \gamma^b_{2,mn} }{\sigma^2_{mn}} \frac{\partial \gamma^b_{2,mn}}{\partial \delta^\mathrm{ic}_p}  \nonumber
    \end{align}

    \begin{align}
       \frac{\partial \log\mathcal{L}} {\partial \delta^\mathrm{ic}_p} &= \sum_b \sum_{mn}  \frac{\hat{\gamma}^b_{1,mn} - \gamma^b_{1,mn} }{\sigma^2_{mn}} \mathrm{DFT}^{-1} \left[ \frac{l_x^2-l_y^2}{l_x^2+l_y^2} \mathrm{DFT}\left(\frac{\partial \kappa_{mn}}{\partial \delta^\mathrm{ic}_p}\right)\right] \nonumber \\ \nonumber
       &+
        \sum_b \sum_{mn}  \frac{\hat{\gamma}^b_{2,mn} - \gamma^b_{2,mn} }{\sigma^2_{mn}} \mathrm{DFT}^{-1} \left[ \frac{-2l_x  l_y}{l_x^2+l_y^2} \mathrm{DFT}\left(\frac{\partial \kappa_{mn}}{\partial \delta^\mathrm{ic}_p}\right)\right]
    \end{align}
    with 
    \begin{align}
       \frac{\partial \kappa_{mn}}{\partial \delta^\mathrm{ic}_p} = \frac{3 H_0^2 \Omega_m}{2 c^2} \Big[\sum\limits_{s=p}^{N_2} \frac{(r_s - r_p)}{r_s} n^b(r_s) \Delta r_s \Big] \frac{r_p \Delta r_p}{a} \frac{\partial \mathcal{M}(a,\delta^\mathrm{ic})}{\partial \delta^\mathrm{ic}_p}
    \end{align}
    where $\mathcal{M}(a,\delta^\mathrm{ic})$ is the dynamical forward model. 
    

\bsp	
\label{lastpage}
\end{document}